# Length-scales of interfacial coupling between metal-insulator phases in oxides

Claribel Domínguez[1*], Alexandru B. Georgescu[2], Bernat Mundet[1,3], Yajun Zhang[4], Jennifer Fowlie[1], Alain Mercy[4], Sara Catalano[1], Duncan T.L. Alexander[3], Philippe Ghosez[4], Antoine Georges[1,2,5,6], Andrew J. Millis[2,7], Marta Gibert[8],

and

Jean-Marc Triscone[1]

[1]Department of Quantum Matter Physics, University of Geneva, Geneva, Switzerland

[2]Center for Computational Quantum Physics, Flatiron Institute, New York, USA

[3]Electron Spectrometry and Microscopy Laboratory (LSME), Institute of Physics (IPHYS), École Polytechnique Fédérale de Lausanne (EPFL), Lausanne, Switzerland

[4]Theoretical Materials Physics, Q-MAT, CESAM, University of Liège, Liège, Belgium

[5]Collège de France, 11 place Marcelin Berthelot, 75005 Paris, France

[6]Centre de Physique Théorique (CPHT), CNRS, Institut Polytechnique de Paris, France

[7]Department of Physics, Columbia University, New York, USA

[8]Physik-Institut, University of Zurich, Winterthurerstrasse 190, 8057 Zurich, Switzerland

**Controlling phase transitions in transition metal oxides remains a central feature of both technological and fundamental scientific relevance. A well-known example is the metal-insulator transition which has been shown to be highly controllable while a less well understood aspect of this phenomenon is the length scale over which the phases can be established. To gain further insight into this issue, we have atomically engineered an artificially phase separated system through fabricating epitaxial superlattices consisting of SmNiO$_3$ and NdNiO$_3$, two materials undergoing a metal-to-insulator transition at different temperatures. By combining advanced experimental techniques and theoretical modeling, we demonstrate that the length scale of the metal-insulator transition is controlled by the balance of the energy cost of the phase boundary between a metal and insulator and the bulk energetics. Notably, we show that the length scale of this effect exceeds that of the physical coupling of structural motifs, introducing a new paradigm for interface-engineering properties that are not available in bulk**.

Phase transitions characterized by a strong coupling of charge, spin, orbital and lattice degrees of freedom are ubiquitous in transition metal oxide (TMO) materials [1]. A well-known example is the metal-insulator transition, which is typically first-order and exhibits regions of phase coexistence. Many studies have focused on manipulating the MIT through intrinsic and extrinsic effects [2-5]. Moreover, experimental techniques such as photoemission electron microscopy and near-field scanning optical microscopy have been used to observe the phase coexistence at the MIT [6,7], however, remarkably little is known about the characteristic lengths of the two possible phases. Important open issues include the minimum length-scale over which a metallic or an insulating state can be established and the physics that sets this length-scale. Is this length-scale controlled by propagation of lattice distortions or more subtle interfacial effects? Answering these questions is important both for understanding the fundamental physics of the metal-insulator transition and for controlling it, essential for application in new generations of electronic devices.

In this paper we present experimental and theoretical analyses of a series of specially-designed structures of two oxide materials that undergo metal-to-insulator transitions at different temperatures while sharing the same order parameter. We create atomic-precision epitaxial superlattices consisting of alternating layers of the two materials and we study the temperature dependence of their conductivity as a function of layer thickness. For thicker layers, two separate transitions are observed such that the engineered bi-component system mirrors the independent properties of the individual layers. For thinner layers the system behaves like one material, with a single metal-insulator transition, implying that we have engineered properties unique to this superlattice system. The characteristic length-scale above which two separate transitions are observed implies a critical minimum phase separation size of around 3 nm. This is remarkably large considering the first-order nature of the metal-insulator transition which is not associated with diverging length scales. By combining our data with theoretical modeling, we demonstrate that the length-scale of the metal-insulator transition in the superlattices is set neither by the length-scale of the propagation of structural motifs nor by that of the electronic order parameter from one material to the other. Rather, the length scale is set by the interplay of phase boundary and bulk energetics. This work illustrates how atomically-engineered layered materials provide a new path for understanding fundamental physics and for engineering functional properties that are not available in bulk.

As a model system for this study we take $NdNiO_3$ and $SmNiO_3$, two members of the rare earth nickelate family $RNiO_3$, where R is a rare earth cation. The rare earth nickelate materials are well known for the sharp metal-to-insulator transition (MIT), observed when temperature is decreased below a characteristic temperature $T_{MI}$, which depends on the choice of rare earth ion R [8-11]. The $T_{MI}$ for bulk $SmNiO_3$ is ~400 K; and for bulk $NdNiO_3$, ~200 K. The rare earth nickelates crystallize in slightly distorted versions of the perovskite structure, which can be visualized as a network of corner-sharing $NiO_6$ octahedra with the rare earth ions occupying the spaces in between. The distortions involve rotations and tilts of the octahedra, which are controlled by the ionic radius of the rare earth ions. The rotations and tilts modify the Ni-O-Ni bond angle ($\Phi$), which sensitively determines the bandwidth of the system and therefore $T_{MI}$ ($\Phi_{NdNiO_3} \approx 157°$ and $\Phi_{SmNiO_3} \approx 153°$ at room temperature in bulk) [9,12-14]. The MIT involves a two-sublattice modulation of electron density (often referred to as "charge order" although the terms "bond density wave" or "charge density disproportionation" as well as "site-selective Mott transition" may be more appropriate [15]) accompanied by a two-sublattice breathing distortion of the $NiO_6$ octahedra. However, understanding and controlling the length-scale over which an MIT can develop is a less well-understood aspect of this phenomenon.

To address these questions, epitaxial superlattices consisting of $L$ repetitions of the basic (($SmNiO_3$)$_m$/($NdNiO_3$)$_m$) unit ($m$ unit cells (u.c.) of $SmNiO_3$ and $m$ u.c. of $NdNiO_3$), were deposited on top of $(001)_{pc}$-oriented $LaAlO_3$ substrates. The total thickness of all the samples was kept at ~40 nm. Therefore, in order to match this thickness, the number of repetitions, *L*, was adjusted for each $((SmNiO_3)_m/(NdNiO_3)_m)$ unit or superlattice wavelength $\Lambda$ (in u.c.) equal to *2m* (*m* in pseudocubic (pc) u.c.), see Fig. 1a). The superlattices were characterized ex-situ by X-ray diffraction, atomic force microscopy and aberration-corrected scanning transmission electron microscopy in combination with electron energy-loss spectroscopy (STEM-EELS). As shown in Fig. 1b), finite thickness oscillations and satellite peaks corresponding to the superlattice periodicity are observed, indicating high quality samples and good agreement with the designed layering. The atomically-flat surface reveals that the superlattices preserve the step-like topography of the substrate (see the corresponding inset in Fig. 1b). Reciprocal space mapping confirms that all superlattices are coherently strained to the substrate (see Supplementary Fig. S1). The structural properties and the chemical composition of the designed $(m,m)_L$ superlattices have been evaluated in detail by means of STEM-EELS. Fig. 1c) shows a gray-scale high angle annular dark field (HAADF), i.e. Z-contrast, image of a $(5,5)_{10}$ superlattice viewed along the [100] zone axis of the $LaAlO_3$ substrate. The HAADF image confirms that the superlattice is fully epitaxial with the substrate and that no obvious defects such as dislocations or Ruddlesden-Popper faults are generated, either within the crystal or at the interfaces. EELS compositional maps obtained from the area indicated on the HAADF image are displayed in panel d) where the La, Sm, Nd, and Ni signal intensities are represented in orange, red, green and blue respectively. All the imaged interfaces are atomically sharp and no obvious cationic intermixing between the $NdNiO_3$ and $SmNiO_3$ layers is identified. The same crystal texture and interface quality have been observed in a $(13,13)_4$ superlattice.

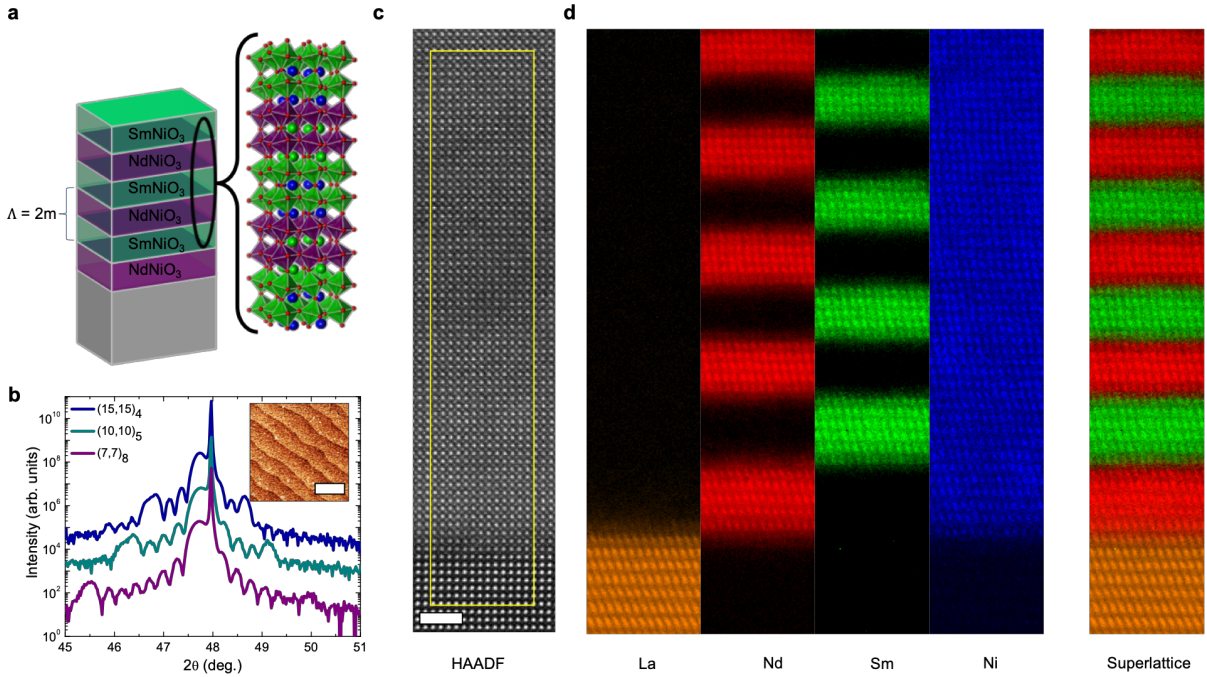

**Figure 1 | Detailed characterization of the (*m*SmNiO$_3$/*m*NdNiO$_3$)$_L$ superlattices. a**, Schematic representation of the superlattice heterostructures, $\Lambda$ indicates the superlattice period. **b**, X-ray diffractograms for (7,7)$_8$, (10,10)$_5$ and (15,15)$_4$ superlattices. The satellite peaks indicate the periodicity of the superlattices. Inset: Typical atomic force microscope topography of a (7,7)$_8$ superlattice. Scale bar: 500 nm. **c**, Cross sectional HAADF image of a (5,5)$_{10}$ superlattice viewed along the [100] zone axis direction of the LaAlO$_3$ substrate. The scale bar corresponds to 2 nm. **d**, EELS compositional maps obtained from the area indicated in (c). The La, Sm, Nd and Ni signals are represented in orange, red, green and blue colors, respectively. The La, Nd and Sm signals are displayed together in the superlattice map.

Fig. 2 displays the resistivity as a function of temperature for three SmNiO$_3$/NdNiO$_3$ superlattices with different periods $\Lambda$. The dashed lines indicate the T$_{MI}$ of 10 nm thick

NdNiO$_3$ and SmNiO$_3$ films grown on (001)$_{pc}$-oriented LaAlO$_3$. As can be seen in Fig. 2, superlattices with large $\Lambda$ ((10,10)$_5$ and (15,15)$_4$), i.e. thick SmNiO$_3$ and NdNiO$_3$ layers, display two distinct MITs. The lower MIT temperature, $T_L$, seen at 100 K < $T_L$ < 200 K is accompanied by a hysteresis loop and resembles the MIT behavior observed in both bulk NdNiO$_3$ ($T_{MI}$ ~ 200 K) and (001)$_{pc}$-oriented epitaxial NdNiO$_3$ thin films ($T_{MI}$ ~ 100 K, for a 10 nm NdNiO$_3$ film deposited on top of LaAlO$_3$ substrate)[16]. The second MIT temperature, $T_H$, manifests at higher temperatures, 300 K < $T_H$ < 400 K, and therefore resembles the MIT of SmNiO$_3$. Interestingly, superlattices with short $\Lambda$ (see for instance (7,7)$_8$) appear to go from a fully metallic to a fully insulating state through a single MIT. By analyzing the entire data set (see Supplementary Fig. S2), we see that the conductive behavior of the higher period SmNiO$_3$/NdNiO$_3$ superlattices tends toward the independent SmNiO$_3$ and NdNiO$_3$ single film behaviors. With decreasing $\Lambda$, the two MITs come closer together in temperature until a joint MIT is measured at a temperature of $T_J$ ~ 155 K. This behavior with a unique MIT is observed below a critical wavelength, $\Lambda_c$, of 16 u.c. These results are summarized on Fig. 4a), to which we will return shortly.

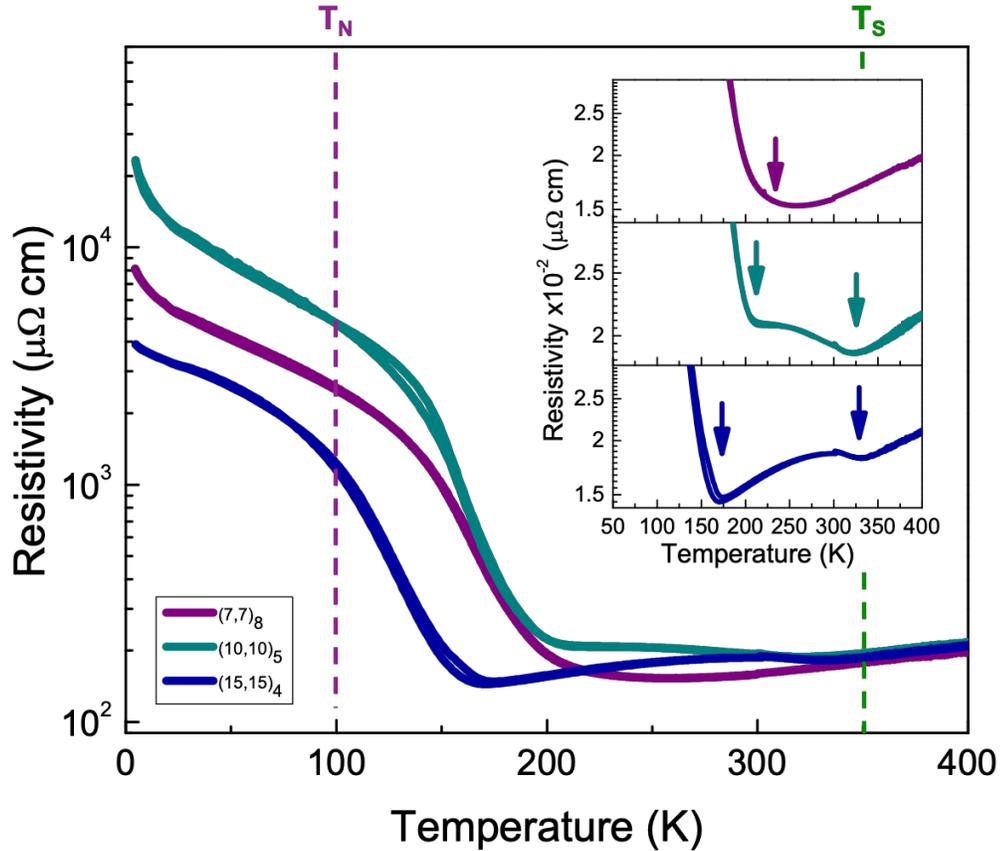

**Figure 2 | Transport measurements.** Resistivity vs temperature of the $(7,7)_8$, $(10,10)_5$ and $(15,15)_4$ $(001)_{pc}$-oriented SmNiO$_3$/NdNiO$_3$ superlattices. The dashed lines indicate the MIT temperatures of NdNiO$_3$ ($T_N$) and SmNiO$_3$ ($T_S$) films of 10 nm thickness on $(001)_{pc}$-oriented LaAlO$_3$ substrates to be compared with the values obtained for the superlattices. Inset: Close ups of the transitions.

The resistivity trends clearly demonstrate that a remarkably long length-scale of 8 u.c. (~3 nm) $\left(\frac{\Lambda_c}{2}\right)$ is required to decouple the two materials and to establish independent phases determined by the bulk energetics. We now consider possible explanations for this long length-scale. As a first step we address the issue of cationic intermixing. We observe that a solid solution $Sm_{0.5}Nd_{0.5}NiO_3$ exhibits a single MIT at approximately the same temperature as that of the short-wavelength superlattices, see Supplementary information Fig. S2. However, as can be seen from Fig. 1d, STEM analyses performed on a $(5,5)_{10}$ superlattice show that intermixing is extremely limited in these structures and does not extend further than 1 u.c. from the interfaces. We can therefore safely say that the long coupling length-scale is not chemical in origin.

A second possibility is interfacial structural coupling, shown in previous studies to be important in many perovskite superlattices [17-20]. In particular, several works have found that rotations and tilts of the oxygen octahedral cages, determining the B-O-B bond angle in $ABO_3$ perovskites, can be modulated across interfaces [21-25]. We also consider for this specific nickelate system that the breathing distortion (bond disproportionation) could gradually propagate from one material to the other. In order to investigate interfacial structural coupling, we determined the crystal structure of $SmNiO_3/NdNiO_3$ superlattices from density functional theory (DFT) + $U$ calculations and corroborated our findings using STEM analyses that allowed the oxygen positions to be resolved. In the calculations, we considered the low temperature phases of $SmNiO_3$ and $NdNiO_3$, which have $P2_1/n$ symmetry (with Glazer notation $a^-b^+a^-$), and the in-phase rotation axis has been oriented in-plane as predicted from the calculations and observed in our STEM measurements. Further details are reported in the methods section and Supplementary information.

Fig. 3a) displays the calculated layer-resolved (z-direction) equatorial and apical Ni-O-Ni bond angle $\Phi$, two equatorial angles coming from the low symmetry. We see that $\Phi$ reverts to a value characteristic of the given nickelate block within only 1 Ni layer of the interface. The STEM imaging is now used to compare the real structure of our $RNiO_3$ superlattices with those simulated by DFT, and to corroborate the predicted short-range coupling of the Ni-O-Ni bond angle at the $SmNiO_3/NdNiO_3$ interfaces. In order to evaluate how these rotations vary through the film depth (z-direction), we imaged the films along two different orthorhombic (ortho) zone axes: the $[110]_{ortho}$ and $[001]_{ortho}$. The former allows evaluation of the rare earth anti-polar shifts ($\Delta Z$) typical of *Pbnm* structures, featuring a zig-zag pattern along the $[001]_{ortho}$ direction, which in itself is an indirect signature of the amplitude of octahedral tilting [24,26]. The latter, $[001]_{ortho}$ zone axis is instead the optimum orientation for measuring $\Phi$, as the oxygen octahedra only rotate in phase along the $[001]_{ortho}$ direction, and therefore all the oxygen atoms belonging to the same atomic column are projected at the same position of the image. Imaging is done using the HAADF and annular bright field imaging (ABF) detectors simultaneously in order to identify the oxygen columns as well as the cation positions. Fig. 3c) shows HAADF (left) and contrast-inverted ABF (right) images obtained from a central region of the $(5,5)_{10}$ superlattice sample. Each image was acquired from equivalent but distinct sample regions, along their corresponding zone axes. An enlarged view of the contrast-inverted ABF image is displayed in Fig. 3d). The illustration includes an overlay of the projection of the corresponding DFT-simulated lattice, which matches very well with the experimental structure. From the HAADF and contrast-inverted ABF images, we have quantified both the $\Delta Z$ and $\Phi$ structural parameters across the superlattice layers, as plotted in Fig. 3f) and g) as red and blue depth-profiles respectively. Both parameters evolve similarly along the z-direction: switching from a specific $NdNiO_3$ value (27 pm and 163º) to another for the $SmNiO_3$ layer (40 pm and 167º), and back again, over only 1 u.c.

from the SmNiO$_3$/NdNiO$_3$ interfaces, thus corroborating the ab-initio calculations. Furthermore, Fig. 3e) shows the DFT-simulated Ni-O-Ni bond angle values projected onto the a-c plane, as this effectively reproduces the geometry of the STEM measurements. Comparing with Fig. 3g) we see the absolute values agree within 3º and the z-dependence is very similar, recovering rapidly to the specific layer value within 1 u.c. We therefore disregard bond angle coupling as the controlling feature for the heterostructure as a whole.

The breathing distortion yields a bond disproportionation (BD), which can be quantified as the difference between the lengths of the long and short Ni-O bonds ($B_L$ and $B_S$ respectively), $BD = \frac{B_L - B_S}{2}$. In the insulating phase, the BD is slightly greater in SmNiO$_3$ than in NdNiO$_3$, so some coupling of the BD magnitude may be expected. Fig. 3b) illustrates the ab-initio determined BD magnitude across the SmNiO$_3$/NdNiO$_3$ interface. Similarly to the bond angle, it can be seen that this parameter returns to the value characteristic of the given nickelate layer after just 1 u.c. of interface effect.

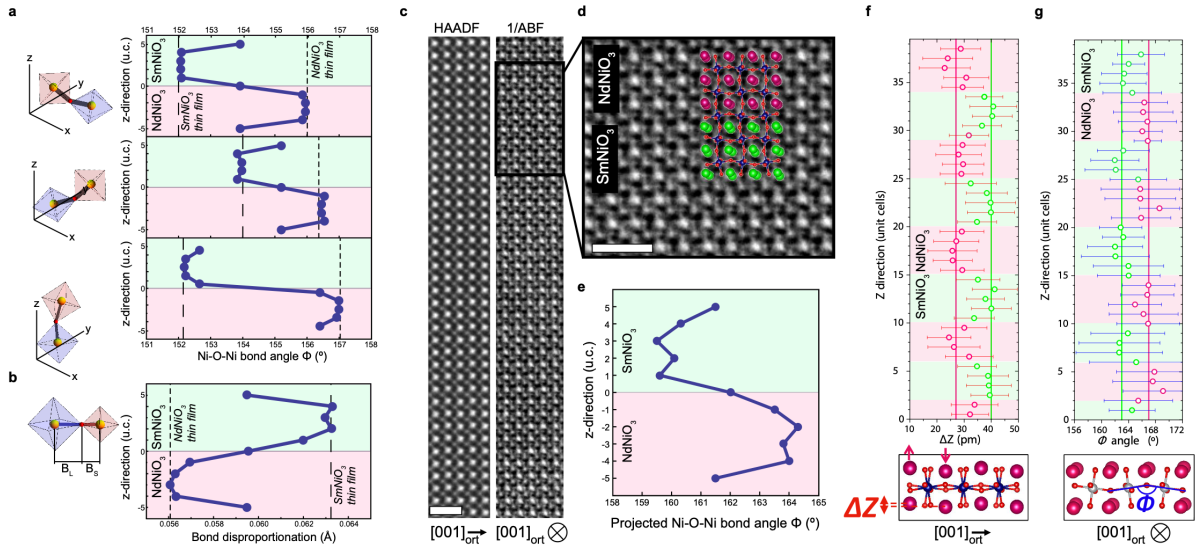

**Figure 3 | DFT calculations performed at 0 K (both materials in the *P2₁/n* phase) of the layer-resolved bond angle and bond disproportionation in the superlattices and STEM structural analyses of the superlattices.** Layer-resolution of the **a,** Equatorial and apical Ni-O-Ni angles **b,** BD magnitude for a (5,5) superlattice as calculated from DFT. Short and long dashed lines indicate the same calculated parameters for thin films of $NdNiO_3$ and $SmNiO_3$ respectively. Schemas are included beside each plot to illustrate the relevant structural parameter. **c,** Cropped areas of a HAADF (left) and a contrast-inverted ABF image (right) obtained from a central region of the experimentally studied $(5,5)_{10}$ superlattice. Scale bar: 1 nm. An enlarged view across a wider region of the contrast-inverted ABF image is shown in **d.** A projection of the $(5,5)_{10}$ DFT-simulated structure is superimposed onto the magnified ABF image. Scale bar: 1 nm. **e,** layer-resolved Ni-O-Ni bond angle projected onto the a-c plane from DFT. **f - g,** $\Delta Z$ (red) and $\Phi$ (blue) depth-profiles obtained from the original HAADF and contrast-inverted ABF images that are shown in c) and d). Both structural parameters are illustrated in the insets. The solid lines represent the corresponding mean values of

these parameters calculated from the interior layers of either the NdNiO$_3$ (magenta) or SmNiO$_3$ (green) block.

From the DFT and the STEM investigations, we conclude that neither bond angle nor bond disproportionation structural motifs propagate beyond 1 u.c. from the interface. This indicates that the effect of purely structural propagation across interfaces cannot account for the persistence of the joint behavior up to a superlattice period of as long as 16 u.c. We propose that the observed behavior is due to the energy cost of establishing a phase boundary – linked to the short-range mismatch of the coupled electronic and structural order parameters corresponding to the metal-insulator transition across the interface. As the structural and electronic order parameters are strongly coupled we can focus on just one. We set the theory in terms of *N* – the electronic disproportionation characterizing the nickelate insulating state. *N* is defined as the difference between the occupation of the long bond and short bond Ni sites ($N = N_{LB} - N_{SB}$) within a low energy, extended antibonding e$_g$ orbital picture [27]. Previous theoretical calculations have found that *N = 0* in the metallic state ($e_g^1$-$e_g^1$) and in the insulating state is around 1.2 - 1.5, depending on the material [27-30].

Based on this electronic disproportionation order parameter we develop a Landau theory, which is described in detail in the Supplementary information. In the simplest description, the key ingredients in the model are the relative free energies of the metallic and insulating phases of the two components of the superlattice, and the energy cost of the phase boundary separating them. Without loss of generality we set the free energy of the metallic phase to zero, $E_M(T) = 0$, and refer to the free energy (per unit cell) of the insulating phase as $E_I(T)$, noting that only the difference between the two is of physical relevance. We take the phase boundary energy between two metallic or two insulating phases to be zero, and the energy of the phase boundary separating the metallic and insulating states to be $E_{PB}$. We assume that the $E_I$ have a linear temperature dependence:

$$E_I^{N,S}(T) = K_{N,S} \frac{(T - T_{N,S})}{T_{N,S}}$$

Here $K$ is the characteristic stiffness of the insulating state in SmNiO3 ($K_S$) and NdNiO3 ($K_N$) and $T_N$ and $T_S$ are the experimental $T_{MI}$s of an individual film of NdNiO3 and SmNiO3 respectively.

We now compute the energy of the three possible states for our (SmNiO3)m/(NdNiO3)m superlattices

1) Both materials are metallic:

$$E_{MM} = mE_M^N + mE_M^S = 0$$

2) NdNiO3 is metallic and SmNiO3 is insulating:

$$E_{MI} = mE_M^N + 2E_{DW} + mE_I^S(T) = 2E_{PB} + mE_I^S(T)$$

3) And both materials are insulating:

$$E_{II} = mE_I^N(T) + mE_I^S(T)$$

Keeping in mind that $\Lambda_c$ ($\Lambda_c = 2m_c$) and $T_J$ are the critical wavelength of the bifurcation and the temperature of the joint $T_{MI}$ coming from experiment respectively and that $m$ designates the number of unit cells of each nickelate block in the repeating unit, one can write three simple equations.

At the bifurcation point, the energies of having both materials insulating ("insulating-insulating"), both metallic ("metallic-metallic") and NdNiO3 metallic and SmNiO3 insulating ("metallic-insulating" phase coexistence with a phase boundary in between costing energy) are equal:

$$m_c E_I^N(T_J) + m_c E_I^S(T_J) = m_c E_M^N(T_J) + m_c E_M^S(T_J) = m_c E_M^N(T_J) + m_c E_I^S(T_J) + 2E_{DW}$$

We may now write an equation for the higher temperature metal-insulator transition ($T_H$), at which $E_{MM}$ becomes equal to $E_{MI}$:

$$m E_M^N(T_H) + m E_M^S(T_H) = m E_M^N(T_H) + m E_I^S(T_H) + 2E_{DW}$$

And the lower transition temperature ($T_L$) at which $E_{II}$ becomes equal to $E_{MI}$:

$$m E_I^N(T_L) + m E_I^S(T_L) = m E_M^N(T_L) + m E_I^S(T_L) + 2E_{DW}$$

Solving this system of equations, the following expressions for $T_L(m)$ and $T_H(m)$ are obtained:

$$T_L = T_N(1 + \frac{2E_{DW}}{mK_N})$$

$$T_H = T_S(1 - \frac{2E_{DW}}{mK_S})$$

For large *m* (large $\Lambda$) the two transitions tend toward the $T_{MI}$s for individual layers of NdNiO3 and SmNiO3 ($T_N$ and $T_S$). When $T_L = T_H$, we obtain the critical wavelength $\Lambda_c$ and $T_J$ (assuming $K_N = K_S$):

$$\frac{\Lambda_C}{2} = m_c = 2E_{PB} \frac{T_S K_N + T_N K_S}{K_S K_N (T_S - T_N)}$$

$$T_J = \frac{2 T_N T_S}{T_N + T_S}$$

The bifurcation point, $\Lambda_c$, is therefore the layer thickness at which the high energy cost of forming a phase boundary between insulating SmNiO$_3$ and metallic NdNiO$_3$ is entirely compensated by the energy gain of the bulk-like phases of the individual layers, making a phase coexisting state, for $\Lambda > \Lambda_c$, energetically feasible.

This simple description provides a qualitative and even semi-quantitative explanation of the behavior observed in our superlattices, as shown by the dotted line in Fig. 4a). Also, $T_J$ is elegantly predicted by this version of the model. Using the experimental values for SmNiO$_3$ and NdNiO$_3$ individual layers ($T_S$ = 352 K and $T_N$ = 100 K), the joint transition temperature ($T_J$) is perfectly estimated at 156 K, in remarkable agreement with experiment.

For the higher temperature branch, however, the fit can be quantitatively improved by refining the model, i.e. introducing a saturation temperature ($T_{Sat}$), releasing the assumption $K_N = K_S$ and using the measured $T_J$ as an input parameter. $T_{Sat}$ restricts the linear temperature dependence of the energy of the SmNiO$_3$ insulating state to a temperature range close to $T_s$ (for $T > T_{Sat}$, $E_I^S(T) = K_S \frac{(T - T_S)}{T_S}$ while for $T < T_{Sat}$, $E_I^S = K_S \frac{(T_{Sat} - T_S)}{T_S}$)). From this input and an estimated value of $T_{Sat}$ we can obtain the following expressions:

$$\frac{K_N}{K_S} = \frac{T_N (T_S - T_{Sat})}{T_S (T_J - T_N)}$$

$$\frac{E_{DW}}{K_N} = \Lambda_C \frac{K_S}{K_N} \frac{(T_S - T_{Sat})}{2T_S}$$

The equations for $T_H$ and $T_L$ remain the same, however, their values are now computed from the newly determined ratios, *$K_N/K_S$ = 0.4* and *$E_{PB}/K_N$ = 4.8.* The excellent fit (dashed line) obtained can be seen in Fig. 4a) that summarizes the experimental data and the output of the Landau theory.

Fig. 4b) and c) display the calculated energies of the three possible phases – insulating-insulating (gray), insulating-metallic (gray/purple) and metallic-metallic (purple) – versus temperature. This is shown for two superlattice wavelengths: $\Lambda$ = 10 u.c. for which, experimentally, the system transitions from a fully metallic state to a fully insulating one and $\Lambda$ = 20 u.c. for which a coexistence of a metallic and an insulating phase is found at intermediate temperatures. As can be seen in Fig. 4b), in the case of $\Lambda$ = 10 u.c., the lowest energy is the metallic-metallic phase at high temperature with a transition to an insulating-insulating phase at 160 K. Examining the calculations for the $\Lambda$ = 20 u.c. case (Fig. 4c)), two transitions are seen to occur: a first one from a high temperature metallic-metallic phase to a mixed (metallic-insulating) phase and, at a lower temperature a second one from the mixed (metallic-insulating) phase to the insulating-insulating one – neatly reproducing what is observed experimentally, as shown in Fig. 4a).

Fig. 4d) and e) show the evolution of the order parameter, *N*, across the interface of two different wavelength superlattices, $\Lambda$ = 10 u.c., 20 u.c., for the various possible phases. As shown in Fig. 4e), at an interface between an insulating region and a metallic one, the order parameter has to go from a value of 1.2 to 0 and the energy cost of bending the order parameter goes as $(\nabla N)^2$. It is only in Fig. 4e), for $\Lambda$ = 20 u.c., that the phase coexistence is stable. However, with a reduced superlattice wavelength of 10 u.c. (Fig. 4d)), the higher density of phase boundaries makes the cost of bending the order parameter between the two distinct electronic phases higher than the energy gained by the individual component materials being in their optimal states. Thus, for $\Lambda$ = 10 u.c., phase coexistence is never stable and, over some temperature range, the phase boundary energy forces either SmNiO$_3$ to be metallic or NdNiO$_3$ to be insulating against the energetics of the bulk phases.

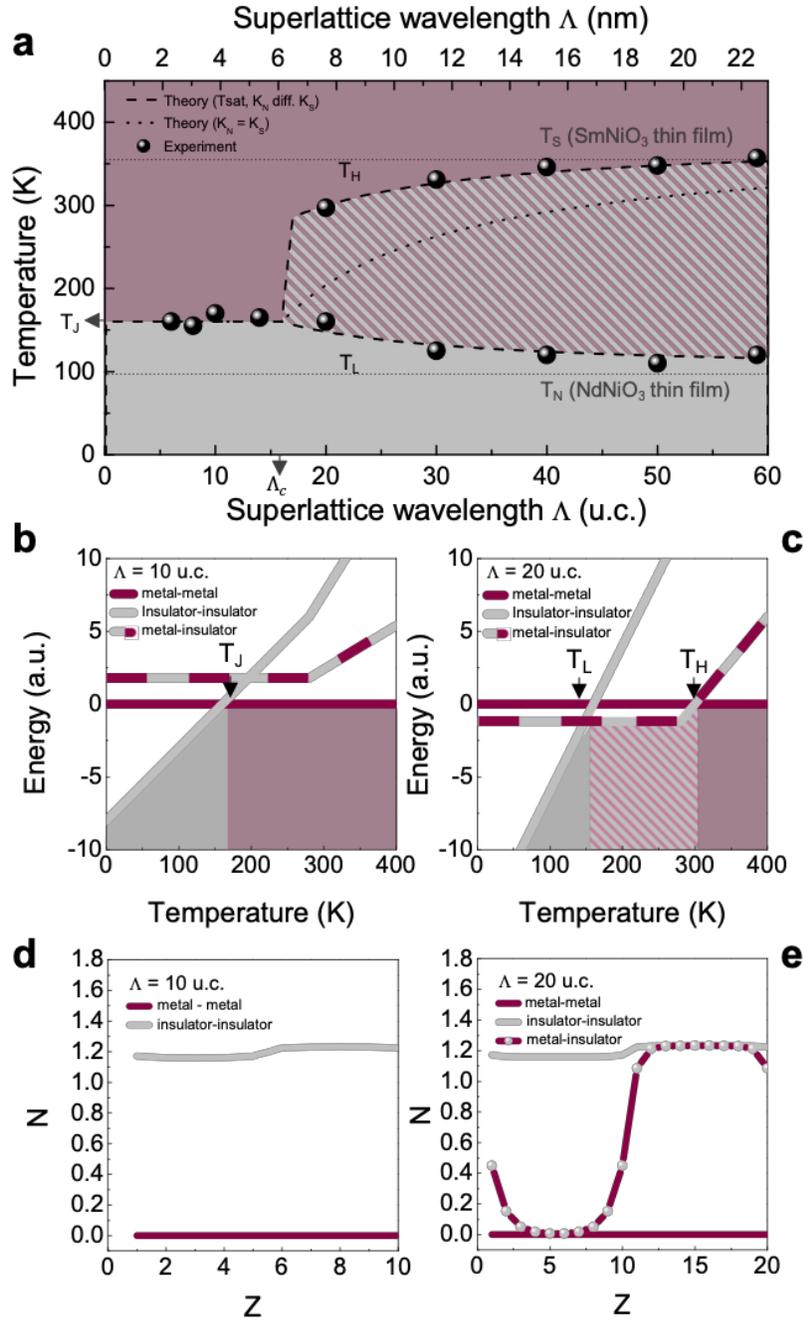

**Figure 4 | Theoretical predictions from the Landau model. a**, Theoretical MIT temperatures (dotted and dashed lines indicate the model results assuming $K_N = K_S$ and introducing $T_{Sat}$ with $K_N \neq K_S$, respectively) as a function of superlattice wavelength and comparison to the experimental data (filled circles). **b - c**, Energy as a function of

temperature for the three possible cases, in two different superlattices ($\Lambda$ = 10 u.c. and $\Lambda$ = 20 u.c.). **d - e**, z-evolution of the order parameter N across the interface for $\Lambda$ = 10 u.c. and $\Lambda$ = 20 u.c. for the different possible cases.

This critical wavelength of 16 u.c is now readily understood despite being astonishingly-long given the short-range nature of the structural motifs, and of the order parameter itself. As we have demonstrated, the Landau expansion based on a first-order transition describes the phase separation size dependence of the phase transition in this superlattice system qualitatively and quantitatively. The ratio between the phase boundary energy and the free energy cost of the thin films to be in the unfavored state is thus the main parameter that determines the physics of the system.

Through engineering specifically-designed superlattices, we were able to clearly isolate the role of the different physical phenomena in setting the length-scale of the MIT in the nickelates. We found that the transition length-scale is controlled by an interfacial phase boundary energy. Notably, it is not determined by long-range propagation of bond disproportionation nor by the direct long-range physical coupling of bond angles, which up until now has been the standard approach to understanding interfacially-coupled behavior. This result introduces a new paradigm for engineering functional properties in heterostructures and can be applied to other combinations of materials that exhibit similar order parameters, as well as to the naturally occurring phase coexistence at a first-order phase transition. Further, we have shown that heterostructures can be used not only to engineer new properties, but as a model system to provide new insights into the fundamental physics of the bulk materials themselves.

## Methods

### Sample preparation

We used radiofrequency off-axis magnetron sputtering at a temperature of 460 °C and a pressure of 0.18 Torr (Ar:O$_2$ mix of ratio 3:1) to grow epitaxial superlattices consisting of $L$ repetitions of the basic ((SmNiO$_3$)$_m$/(NdNiO$_3$)$_m$) unit ($m$ unit cells of SmNiO$_3$ and $m$ unit cells of NdNiO$_3$) on top of (001)$_{pc}$-oriented LaAlO$_3$ substrates. Prior to deposition, the LaAlO$_3$ substrates were thermally treated to ensure atomically flat terraces and step-like topography. Bulk SmNiO$_3$ and NdNiO$_3$ have pseudocubic lattice constants of $a_{pc}$ = 3.799 Å and $a_{pc}$ = 3.808 Å respectively; whereas $a_{pc}$ = 3.787 Å for LaAlO$_3$. Hence the corresponding lattice mismatch $(a_{substrate}-a_{film})/a_{film}$ to the substrate is $\varepsilon_{xx}$ = - 0.3 % and $\varepsilon_{xx}$ = - 0.6 % for the SmNiO$_3$ and NdNiO$_3$ layers, respectively.

### Transport measurements

Transport measurements were carried out in a 4 points configuration, after being patterned with UV lithography in standard Hall bar geometry with Pt contacts, in the temperature range 4 K < T < 400 K. Below 300 K, the samples were slowly dipped into a liquid helium bath while higher-than-room temperature measurements were achieved using two Peltier elements. The $T_{MI}$ of the superlattices was determined from the maximum of $-d(lnR)/dT$ on heating [31].

### Aberration-corrected STEM measurements

STEM specimens were prepared by mechanical tripod polishing, followed by argon ion beam milling to electron transparency with a Gatan PIPS II. Simultaneous series of high angle annular dark field (HAADF) and annular bright field (ABF) STEM images were acquired using a double-aberration-corrected Titan Themis 60-300 microscope located at the CIME (EPFL). The instrument was operated at 300 keV, using an ~20 mrad convergence semi-angle for the electron probe. From each image series, the Smart Align software was used to produce an averaged image with reduced statistical image noise and correction of linear and non-linear scan distortions [32]. The central positions of all the imaged atomic columns were identified by using an iterated refinement process, which ensures a subatomic precision that localizes the center of mass position associated to each atomic column. From the obtained set of coordinates, one per atomic sub-lattice, we can estimate the depth-evolution of any structural parameter. Each data point shown in Fig. 3f) and g) corresponds to the mean over 30 u.c. and the error bars derive from the standard deviation.

STEM-EELS spectrum image (SI) datasets were acquired with a Gatan GIF Quantum ERS spectrometer, using the following conditions: a collection semi-angle of ~47 mrad; 0.01 s acquisition time per pixel; 0.5 eV/ch energy dispersion and 2048 channels, thus covering all the spectral edges located between 400 – 1424 eV in energy loss. Principal component analysis was used to improve the signal-to-noise ratio of the EELS-SI datasets using a plugin in Gatan DigitalMicrograph (DM) [33,34]. In particular, we used the first 14 principal components to reconstruct the EELS-SI for the maps shown in Fig. 1c). The *La* $M_{45}$, the *Ni* $L_{23}$, *Sm* $M_{45}$ and *Nd* $M_{45}$ edges were used for mapping the relative composition of the superlattices. A power-law dependency was used to fit and subtract the background signal from the edges. The multiple linear least squares plugin within DM was used to deconvolute the *La* $M_{45}$ and *Ni* $L_{23}$ peak signals, as they are substantially overlapped.

**DFT**

Our first-principles calculations were carried out using density functional theory (DFT) with Projector Augmented Wave (PAW) method as implemented in the Vienna ab initio simulation package (VASP) [35-37]. We used the PBEsol [38] exchange-correlation energy functional. In line with recent work on RNiO$_3$ compounds [39,40], an on-site Coulomb interaction [41] $U = 2\ eV$ was added to the 3$d$ orbitals of Ni according to the method of Dudarev [42]. The plane-wave energy cutoff was set to 700 eV. The (NdNiO$_3$)$_1$/(SmNiO$_3$)$_1$, (NdNiO$_3$)$_3$/(SmNiO$_3$)$_3$ and (NdNiO$_3$)$_5$/(SmNiO$_3$)$_5$ superlattices were simulated respectively using 2 × 2 × 10, 2 × 2 × 6 and 2 × 2 × 10 supercells and 6 × 6 × 1, 6 × 6 × 2 and 6 × 6 × 1 Monkhorst-Pack k-point meshes [43]. The lattice constants of fully relaxed bulk NdNiO$_3$ and SmNiO$_3$ are in good agreement with experimental data (see Supplementary Table S1). In order to simulate properly the epitaxial strain imposed experimentally to NdNiO$_3$ and SmNiO$_3$ on a LaAlO$_3$ substrate, the in-plane lattice parameters of the superlattices were fixed to be 3.77 Å. The oxygen rotation pattern was oriented with the long axis (c$^+$) in-plane (see Supplementary information). The out-of-plane lattice parameters and internal atomic coordinates were relaxed until the total energy and Hellman-Feynman force were converged to 10$^{-7}$ eV. and 10$^{-3}$ eV/Å respectively. For simplicity, all the calculations were performed in the ferromagnetic state. Bond lengths and bond angles were calculated using the VESTA package [44].

**Data Availability**

The data that support the findings of this study are available from the corresponding author upon reasonable request.

**Code Availability**

The computer codes and algorithm used to generate results that are reported in the paper are available from the corresponding author upon reasonable request.

**References**

1       Khomskii, D. I. *Transition Metal Compounds*.  (Cambridge University Press, 2014).

2       Lee, D. *et al.* Isostructural metal-insulator transition in $VO_2$. *Science* **362**, 1037-1040, (2018).

3       Cen, C. *et al.* Nanoscale control of an interfacial metal-insulator transition at room temperature. *Nat. Mater.* **7**, 298-302, (2008).

4       Thiel, S., Hammerl, G., Schmehl, A., Schneider, C. W. & Mannhart, J. Tunable Quasi–Two-Dimensional Electron Gases in Oxide Heterostructures. *Science* **313**, 1942-1945, (2006).

5       Zubko, P., Gariglio, S., Gabay, M., Ghosez, P. & Triscone, J.-M. Interface Physics in Complex Oxide Heterostructures. *Annu. Rev. Condens. Matter Phys.* **2**, 141-165, (2011).

6       Mattoni, G. *et al.* Striped nanoscale phase separation at the metal-insulator transition of heteroepitaxial nickelates. *Nat. Commun.* **7**, 13141, (2016).

7       Post, K. W. *et al.* Coexisting first- and second-order electronic phase transitions in a correlated oxide. *Nat. Phys.* **14**, 1056–1061, (2018).

8       Lacorre, P. *et al.* Synthesis, crystal structure, and properties of metallic $PrNiO_3$: Comparison with metallic $NdNiO_3$ and semiconducting $SmNiO_3$ *J. Solid State Chem.* **91**, 225-237, (1991 ).

9   Torrance, J. B., Lacorre, P., Nazzal, A. I., Ansaldo, E. J. & Niedermayer, C. Systematic study of insulator-metal transitions in perovskites RNiO$_3$ (R = Pr, Nd, Sm, Eu) due to closing of charge-transfer gap. *Phys. Rev. B* **45**, 8209-8212, (1992).

10  Middey, S. *et al.* Physics of ultrathin hilms and heterostructures of rare-earth nickelates. *Annu. Rev. Mater. Res.* **46**, 305-334, (2016).

11  Catalano, S. *et al.* Rare-Earth Nickelates RNiO$_3$: Thin films and heterostructures. *Rep. Prog. Phys.* **81**, (2017).

12  Medarde, M. L. Structural, magnetic and electronic properties of RNiO$_3$ perovskites (R = rare earth). *J. Phys. Condens. Matter* **9**, 1679–1707, (1997).

13  Catalan, G. Progress in perovskite nickelate research. *Phase Transit.* **81**, 729-749, (2008).

14  Mercy, A., Bieder, J., Iniguez, J. & Ghosez, P. Structurally triggered metal-insulator transition in rare-earth nickelates. *Nat. Commun.* **8**, 1677, (2017).

15  Park, H., Millis, A. J. & Marianetti, C. A. Site-selective Mott transition in rare-earth-element nickelates. *Phys. Rev. Lett.* **109**, 156402, (2012).

16  Scherwitzl, R. *et al.* Electric-field control of the metal-insulator transition in ultrathin NdNiO$_3$ films. *Adv. Mater.* **22**, 5517-5520, (2010).

17  May, S. J. *et al.* Control of octahedral rotations in (LaNiO$_3$)n/(SrMnO$_3$)m superlattices. *Phys. Rev. B* **83**, 153411, (2011).

18  Rondinelli, J. M., May, S. J. & Freeland, J. W. Control of octahedral connectivity in perovskite oxide heterostructures: An emerging route to multifunctional materials discovery. *MRS Bulletin* **37**, 261-270, (2012).

19  Rondinelli, J. M. & Spaldin, N. A. Structure and properties of functional oxide thin films: insights from electronic-structure calculations. *Adv. Mater.* **23**, 3363-3381, (2011).

20  Chakhalian, J., Millis, A. J. & Rondinelli, J. Whither the oxide interface. *Nat. Mater.* **11**, 92-94, (2012).

21  Moon, E. J. *et al.* Spatial control of functional properties via octahedral modulations in complex oxide superlattices. *Nat. Commun.* **5**, 5710, (2014).

22  Aso, R., Kan, D., Shimakawa, Y. & Kurata, H. Atomic level observation of octahedral distortions at the perovskite oxide heterointerface. *Sci. Rep.* **3**, 2214, (2013).

23  Liao, Z. *et al.* Controlled lateral anisotropy in correlated manganite heterostructures by interface-engineered oxygen octahedral coupling. *Nat. Mater.* **15**, 425-431, (2016).

24  Liao, Z. *et al.* Metal-insulator-transition engineering by modulation tilt-control in perovskite nickelates for room temperature optical switching. *Proc. Natl. Acad. Sci. U. S. A.* **115**, 9515-9520, (2018).

25  Yuan, Y. *et al.* Three-dimensional atomic scale electron density reconstruction of octahedral tilt epitaxy in functional perovskites. *Nat. Commun.* **9**, 5220, (2018).

26  Zhang, J. Y., Hwang, J., Raghavan, S. & Stemmer, S. Symmetry lowering in extreme-electron-density perovskite quantum wells. *Phys. Rev. Lett.* **110**, 256401, (2013).

27  Subedi, A., Peil, O. E. & Georges, A. Low-energy description of the metal-insulator transition in the rare-earth nickelates. *Phys. Rev. B* **91**, 075128, (2015).

28   Seth, P. *et al.* Renormalization of effective interactions in a negative charge transfer insulator. *Phys. Rev. B* **96**, 205139, (2017).

29   Peil, O. E., Hampel, A., Ederer, C. & Georges, A. Mechanism and control parameters of the coupled structural and metal-insulator transition in nickelates. *Phys. Rev. B* **99**, 245127, (2019).

30   Georgescu, A. B., Peil, O. E., Disa, A. S., Georges, A. & Millis, A. J. Disentangling lattice and electronic contributions to the metal–insulator transition from bulk vs. layer confined $RNiO_3$. *Proc. Natl. Acad. Sci. U. S. A.* **116**, 14434-14439, (2019).

31   Catalan, G. Metal-insulator transition in $NdNiO_3$ thin films. *Phys. Rev. B* **62**, 7892-7900, (2000).

32   Jones, L. *et al.* Smart Align—a new tool for robust non-rigid registration of scanning microscope data. *Adv. Struc. Chem. Imaging* **1**, (2015).

33   Bosman, M., Watanabe, M., Alexander, D. T. & Keast, V. J. Mapping chemical and bonding information using multivariate analysis of electron energy-loss spectrum images. *Ultramicroscopy* **106**, 1024-1032, (2006).

34   Lucas, G., Burdet, P., Cantoni, M. & Hebert, C. Multivariate statistical analysis as a tool for the segmentation of 3D spectral data. *Micron* **52-53**, 49-56, (2013).

35   Kresse, G. & Furthmüller, J. Efficient iterative schemes for ab initio total-energy calculations using a plane-wave basis set. *Phys. Rev. B* **54**, 11169-11186, (1996).

36   Kresse, G. & Furthmüller, J. Efficiency of ab-initio total energy calculations for metals and semiconductors using a plane-wave basis set. *Comput. Mater. Sci.* **6**, 15-50, (1996).

37   Kresse, G. & Joubert, D. From ultrasoft pseudopotentials to the projector

augmented-wave method. *Phys. Rev. B* **59**, 1758-1775, (1999).

38  Perdew, J. P. *et al.* Restoring the Density-Gradient Expansion for Exchange in Solids and Surfaces. *Phys. Rev. Lett.* **100**, 136406, (2008).

39  Varignon, J., Grisolia, M. N., Íñiguez, J., Barthélémy, A. & Bibes, M. Complete phase diagram of rare-earth nickelates from first-principles. *npj Quantum Materials* **2**, (2017).

40  Hampel, A. & Ederer, C. Interplay between breathing mode distortion and magnetic order in rare-earth nickelates $RNiO_3$ within DFT+U. *Phys. Rev. B* **96**, 165130, (2017).

41  Loschen, C., Carrasco, J., Neyman, K. M. & Illas, F. First-principlesLDA+UandGGA+Ustudy of cerium oxides: Dependence on the effective U parameter. *Phys. Rev. B* **75**, 035115 (2007).

42  Dudarev, S. L., Botton, G. A., Savrasov, S. Y., Humphreys, C. J. & Sutton, A. P. Electron-energy-loss spectra and the structural stability of nickel oxide: An LSDA+U study. *Phys. Rev. B* **57**, 1505-1509, (1998).

43  Monkhorst, H. J. & Pack, J. D. Special points for Brillouin-zone integrations. *Phys. Rev. B* **13**, 5188-5192, (1976).

44  Momma, K. & Izumi, F. VESTA: a three-dimensional visualization system for electronic and structural analysis. *J. Appl. Crystallogr.* **41**, 653-658, (2008).

45  García-Muñoz, J. L., Rodríguez-Carvajal, J., Lacorre, P. & Torrance, J. B. Neutron-diffraction study of $RNiO_3$ (R = La, Pr, Nd, Sm): Electronically induced structural changes across the metal-insulator transition. *Phys. Rev. B* **46**, 4414-4425, (1992).

46   Alonso, J. A., Martínez-Lope, M. J., Casais, M. T., Aranda, M. A. G. & Fernández-Díaz, M. T. Metal−Insulator Transitions, Structural and Microstructural Evolution of RNiO3(R = Sm, Eu, Gd, Dy, Ho, Y) Perovskites: Evidence for Room-Temperature Charge Disproportionation in Monoclinic HoNiO3and YNiO3. *J. Am. Chem. Soc.* **121**, 4754-4762, (1999).

47   Press, W. H., Teukolsky, S. A., Vetterling, W. T. & Flannery, B. P. *Numerical Recipes in C++: The Art of Scientific Computing*. (Cambridge University Press, 2002).

**Supplementary Information**

1. **X-ray reciprocal space mapping of SmNiO$_3$/NdNiO$_3$ superlattices**

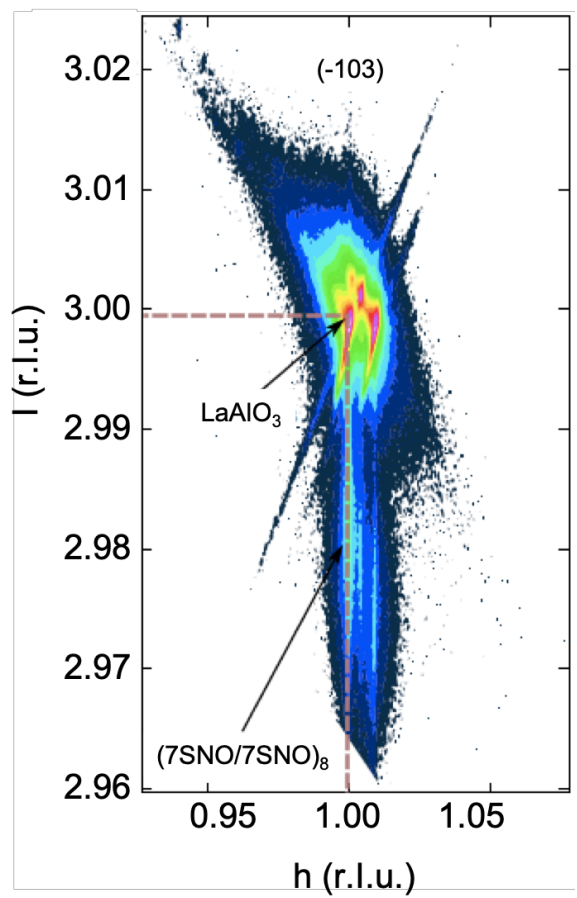

**Supplementary Figure S1 |** X-ray diffraction reciprocal space map for a (7,7)$_8$ superlattice around the (-103) reflection, confirming coherent growth of the superlattice.

## 2. Transport measurements

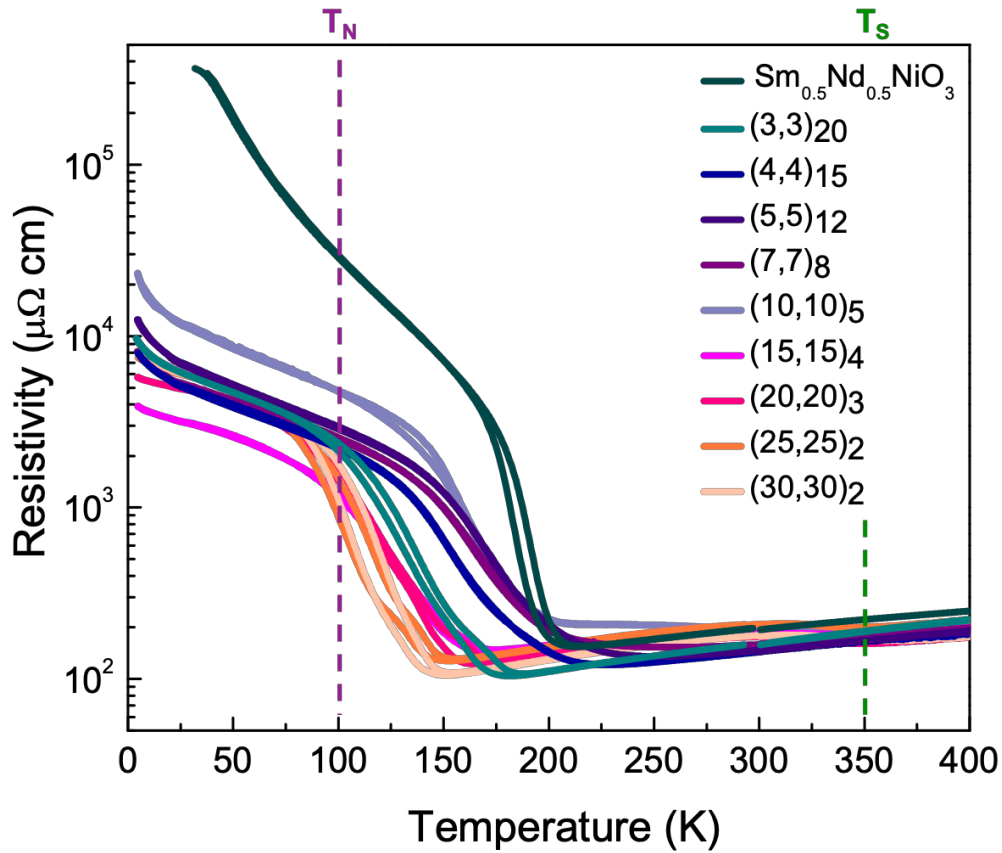

**Supplementary Figure S2 |** Resistivity vs temperature of all $(001)_{pc}$-oriented $SmNiO_3/NdNiO_3$ superlattices with different periods. The dashed lines indicate the MIT temperatures of the corresponding $NdNiO_3$ and $SmNiO_3$ films ~10 nm thick, to be compared with the values obtained for the superlattices.

## 3. Extra text on the structure

Both $SmNiO_3$ and $NdNiO_3$ are slightly distorted perovskite oxides. At higher temperatures both compounds are metallic and orthorhombic with *Pbnm* symmetry, and the rotations and tilts of oxygen octahedra are described by the $a^-a^-c^+$ pattern in Glazer notation. At lower temperatures, both compounds undergo a monoclinic distortion with *P2$_1$/n* symmetry, combining the same pattern of octahedral rotations and tilts with an additional breathing motion of the octahedra [45]. In the Table below, we compare the lattice constants of the bulk *P2$_1$/n* phases, as reported experimentally and computed from DFT. We observe an excellent agreement, with only a systematic trend of the calculations to slightly underestimate the lattice constants (typically by less than 1%). It appears that $SmNiO_3$ has a much larger orthorhombic distortion (*b* > *a*) than $NdNiO_3$, features properly captured by the DFT calculations.

|  | $SmNiO_3$ | | $NdNiO_3$ | |
|---|---|---|---|---|
|  | Experimental[46] | DFT | Experimental[45] | DFT |
| a | 5.327 | 5.258 | 5.386 | 5.370 |
| b | 5.437 | 5.430 | 5.382 | 5.353 |
| c | 7.565 | 7.500 | 7.607 | 7.565 |
| a/√2 | 3.767 | 3.718 | 3.808 | 3.797 |
| b/√2 | 3.845 | 3.840 | 3.806 | 3.785 |
| c/2 | 3.783 | 3.750 | 3.803 | 3.783 |
| $\varepsilon_{zz}$ | +0.12 | +0.53 | -0.43 | -0.33 |
| $\varepsilon_{xx}$ | -0.50 | -0.25 | -0.53 | -0.56 |

**Supplementary Table 1:** Comparison of the experimental and theoretical (ferromagnetic configuration) lattice constants (in Å) of bulk SmNiO$_3$ and NdNiO$_3$ in their *P2$_1$/n* ground-state and related epitaxial strains (in %) on a LaAlO$_3$ substrate for a growth with c axis in-plane.

According to the information provided in the Supplementary Table 1, when growing NdNiO$_3$ and SmNiO$_3$ on top of LaAlO$_3$ ($a_{pc}$ = 3.787 Å), one can expect that the epitaxial strain will not induce any strong preferential orientation for NdNiO$_3$, while it should favor the growth of SmNiO$_3$ with its c axis in-plane. Accordingly, the superlattices should preferentially grow with the same orientation. This is confirmed by the DFT calculations, showing that the configuration with the c-axis in-plane is energetically favored with respect to the configuration with the c-axis out-of-plane. This is also corroborated experimentally by the STEM measurements. In order to reproduce an epitaxial strain state comparable to the experiment (see Supplementary Table 1), we considered in our calculations (that slightly underestimate lattice constants) a pseudo-cubic lattice constant of 3.77 Å for LaAlO$_3$.

## 4. Bulk Landau Theory

In order to study the transition in more detail and thereby justify the analytical approach in the main text, we also perform numerical calculations involving a full, layer dependent bulk Landau theory for the two component materials. The metal-insulator transition in the nickelates is characterized by a simultaneous electronic ($N$) and structural bond disproportionation (BD) from the high temperature metallic phase to the low temperature insulating phase. As one order parameter determines the other, without loss of generality, we select the electronic order parameter $N$ to be the one determining the physical state of the system. As in the main text and previous work [27-30] this is usually defined as the occupancy difference between the higher filling and the lower filling $e_g$ frontier antibonding orbitals in a *d*-orbital only picture.

To model a superlattice of periodicity $\Lambda$ we then define a total number of $\Lambda$ points which we label by coordinate $z$, half of which corresponds to NdNiO$_3$ $\left(1 \leq z \leq \frac{\Lambda}{2}\right)$, and half of which corresponds to SmNiO$_3$ $\left(\frac{\Lambda}{2}+1 \leq z \leq \Lambda\right)$. To each point $z$ we then assign a polynomial form of the free energy. We pick a standard choice for a first-order free energy landscape for each point:

$$E_{bulk}(z,N,T) = P(N) = \frac{A(z)N^6}{6} + \frac{B(z)N^4}{4} + \frac{C(z,T)N^2}{2} \qquad (1)$$

As in the main text, this theory has a free energy of 0 for the metallic phase, independent of temperature at $N(z) = 0$ corresponding to the metallic phase. For the bulk insulating phase, we then use $N(z) = 1.2$ for the NdNiO$_3$ layers and $N(z) = 1.4$ electrons for the insulating state at the metal-insulator transition temperature, consistent with previous DFT+DMFT work [27-30], however noting that these values are not relevant in themselves as we will show later. For each material we then determine the coefficients at the metal-insulator transition temperature of the respective material. We set $A(z) = 1$ for all $z$ (which sets an overall normalization factor for the Landau theory) then determine $B(z)$ and $C(z, T_{MIT})$ from $E(z, N_I, T_{MIT}) = 0$ and $\frac{\partial E_{bulk}(z, N_I, T_{MIT})}{\partial N} = 0$. In order to then tune the free energy to obtain a metal-insulator transition we vary the coefficient $C(T)$ with temperature as: $C(z, T) = C(z, T_{MIT})\left(1 + \alpha(z)\frac{T - T_{MIT}}{T_{MIT}}\right)$ with $0 < \alpha(z) < 1$ and $\alpha(z)$ taking one of two values we determine $\alpha_{NdNiO_3}$ or $\alpha_{SmNiO_3}$. We note that this is equivalent to changing the susceptibility of the metallic phase as $C(z, T) = \chi_M^{-1}$, while keeping $C(z, T)$ in order to ensure the first-order nature of the metal-insulator transition.

To these we add a discretized gradient term that couples nearest neighbor layers by penalizing changes in the order parameter from one layer to the other:

$$E_{gradient} = \sum_z \frac{\xi^2}{2} (N(z) - N(z-1))^2 \qquad (2)$$

where $z$ are the integer values determining the layer number.

The total free energy is then:

$$E_{total} = \sum_z E_{bulk}(z, N, T) + \sum_z \frac{\xi^2}{2}(N(z) - N(z-1))^2 \qquad (3)$$

As the coefficients for the $E(z, N, T)$ are the same throughout each material, but different from one to the next, in the $\xi \to 0$ limit, we obtain a sum of two bulk energies as in the main text. For $\xi \neq 0$, an energy cost will appear from a combination of $N(z)$ not being at the optimum bulk values, and from the order parameter 'bending' across the interface, leading to a phase boundary energy cost. If this phase boundary is restricted in length compared to $\frac{\Lambda}{2}$, then this energy cost will be constant.

We now have effectively three adjustable parameters, $\alpha_{NdNiO_3}$, $\alpha_{SmNiO_3}$ and $\alpha$. Similar to the simplified theory in the main text, the ratio of the two $\alpha$ can be obtained from the joint transition temperature at the bifurcation point $T_J$ (similar to the $K$ constants in the text), while $\xi$ can be obtained from the critical wavelength of the bifurcation point $\Lambda_c$. Note that one can adjust the $N_I$ for the insulating states of the materials as well, however as long as the $\alpha$ and $\xi$ are determined as described above after picking the $N_I$, the resulting transition temperatures will be the same (confirming that the main physics is in fact captured by the simplified model in the main text).

The final step is to initialize and solve the calculation for the three physically relevant configurations. To obtain a full metallic solution, we initialize the calculation with $N(z) = 0$ for all $z$, for a mixed NdNiO3 metallic, SmNiO3 insulating solution we initialize $N(z) = 0$ for $1 \leq z \leq \frac{\Lambda}{2}$ and $N(z) = N_I^{SNO}$ for $\frac{\Lambda}{2} + 1 \leq z \leq \Lambda$ and for the full insulating solution we initialize $N(z) = N_I$ with the $N_I$ corresponding to the respective material. Then we minimize over the total free energy and depending on which initial state we started in we obtain a different minimum corresponding to one of the three possible solutions. Similar to the simplified theory, we can additionally impose a saturation temperature for $C(z,T)$ for the SmNiO$_3$ layers: $C(z,T) = C(z,T_{MIT})\left(1 + \alpha(z)\frac{T_{saturation}-T_{MIT}}{T_{MIT}}\right)$ for $T < T_{saturation}$.

### 5. About Landau Model

We find minima of the polynomial Landau theory numerically using a standard leapfrog scheme which is iterated to self-consistency. Calculations are initialized using all possible combinations of the solutions for bulk phases, thereby producing all insulating, all metallic and mixed solutions [47]

**Acknowledgements**

We thank Hugo Strand and Manuel Zingl for fruitful discussions and acknowledge Marco Lopes and Sébastien Muller for their invaluable technical support. This work was partly supported by the Swiss National Science Foundation through Division II. The research leading to these results has received funding from the European Research Council under the European Union's Seventh Framework Program (FP7/2007-2013)/ERC Grant Agreement 319286 Q-MAC). The authors acknowledge access to the electron microscopy facilities at the Interdisciplinary Centre for Electron Microscopy (CIME), EPFL. The Flatiron Institute is a division of the Simons Foundation. Ph.G., Y.Z. and A.M. acknowledge support from ARC project AIMED and M-ERA.NET project SIOX as well as access to computational resources provided by the Consortium des Equipements de Calcul Intensif (CECI), funded by the Belgian F.R.S.-FNRS under the Grant No. 2.5020.11 and the Tier-1 supercomputer of the Fédération Wallonie-Bruxelles funded by the Walloon Region of Belgium under the Grant No 1117545. M.G acknowledges support by the Swiss National Science Foundation under grant No. PP00P2_170564.

**Author contributions**

M.G. and J-M.T. conceived the project. C.D. fabricated the superlattices and carried out the transport measurements. The Landau model was developed by A.G., A.G. and A.M. TEM was performed and analyzed by B.M. and D.A. The first-principles calculations were carried out by Y.Z., A.M. and Ph.G. C.D. and J.F wrote the manuscript with input from all authors. All authors contributed to the analysis and interpretation of the experimental results.